# Identification of Group Changes in Blogosphere

Bogdan Gliwa[1], Stanisław Saganowski[2], Anna Zygmunt[1], Piotr Bródka[2], Przemysław Kazienko[2], Jarosław Koźlak[1]

[1]AGH University of Science and Technology, Al. Mickiewicza 30, 30-059 Kraków, Poland
[2] Institute of Informatics, Wrocław University of Technology, Wyb.Wyspiańskiego 27, 50-370 Wrocław, Poland
bgliwa@agh.edu.pl, stanislaw.saganowski@pwr.wroc.pl , azygmunt@agh.edu.pl, piotr.brodka@pwr.wroc.pl, kazienko@pwr.wroc.pl, kozlak@agh.edu.pl

*Abstract*—**The paper addresses a problem of change identification in social group evolution. A new SGCI method for discovering of stable groups was proposed and compared with existing GED method. The experimental studies on a Polish blogosphere service revealed that both methods are able to identify similar evolution events even though both use different concepts. Some differences were demonstrated as well.**

*Keywords - group identification, group changes, blogoshpere, social community, dynamics analysis*

I. INTRODUCTION RELATED WORK

A social group is a set of social entities (people or groups of people) closely interlinked and simultaneously relatively far from other groups. This corresponds to groups of friends, families or project teams usually having a strong sense of community. It is equivalent to groups of people involved in the same kind of activities (common interest) in the virtual world, especially in the blogosphere. It means that users posting and commenting on each other are driven by similar topics and form a kind of special interest groups (SIG). On the other hand these virtual communities evolve and move through various stages – a sequence of consecutive events. Identification of changes in social groups evolution is the main goal of this paper.

*A. Group extraction*

There are many definitions of groups, mainly according to the area in which they were created. For example, according to social identity theory [21], a "social group consists of a number of people who feel and perceive themselves as belonging to this group and who are said to be in group by others". Most definitions arising in sociology are very similar, but on the basis of them, it would be difficult to extract measurable characteristics. In addition, virtual communities (which can be extracted from social media e.g. blogs) differ from physical ones. In [6], Chin and Chignel tried do adopt features of a sense of community (such as: feeling of membership, feeling of influence, reinforcement of needs and shared emotional connections) to their method of identifying and measuring virtual communities in blogs. The idea of finding groups in a social network is to identify a set of vertices, which communicate to each other more frequently than with vertices outside the group [9], [22], [23]. For example, in blogosphere, groups are not isolated, but individuals can, in a given time, be a member of many groups

Many methods of finding groups (overlapping and not) have been proposed. In [8] there are detailed descriptions of the most popular methods and algorithms.

*B. Group evolution*

The majority of works on finding groups in the community assume that the graph is static [9], [7]. This is an unrealistic assumption especially in the case of social media (e.g. Facebook, YouTube, forums, blogosphere) which are very dynamic and which evolve over time. Therefore, a growing interest in developing algorithms for extracting communities that take into account the dynamic aspect of the network has been observed for some time. For example Palla [18], [19] has expanded its classical algorithm CPM (Clique Percolation Method) finding k-clique [16], [17]. In [14] authors analyse network dynamics based on different topological properties (e.g degree distribution, small world properties). Similarly in [3], they examine how the evolution of groups relates to structural properties of the network. In [23] a method for tracking groups by adaptive evolutionary clustering with additional temporal smoothing is proposed.

A method of tracking groups over time, regardless of which way they were extracted, was proposed in [10]. First, a division into time steps is carried out. At each step, the graph is created and groups are extracted. Groups of consecutive time steps are adjusted, using the Jaccard index and define a threshold above which the continuation of the group is assumed.

Backstrom [3] analysed group formation focusing especially on group membership, growth and change. Palla et al. in [18] identified basic events that may occur in the life cycle of the group: growth, merging, birth, construction, splitting and death. They didn't give any additional conditions. Asur in [2]introduced formal definitions to five critical events. Green in [10] found that it exists rather consensus on the fundamental events describing group evolution and formulated these key events in terms of rules.

In recent years, several other methods for tracking changes in social groups have been proposed. Sun et al. have introduced GraphScope [20], Chakrabarti et al. have presented another original approach in [5], Lin et al. have provided the framework called FacetNet [15] using evolutionary clustering, Kim and Han in [13] have introduced the concept of nano-communities, Hopcroft et al. have also investigated group evolution, but no method which can be implemented have been provided [11]. This variety of methods and approaches suggests that group evolution is very important research problem.

II. SGCI: THE ALGORITHM FOR STABLE GROUP CHANGES IDENTIFICATION

The used model of the blogosphere takes into consideration its dynamic behaviour and is prepared for the

The work was partly supported by The Polish National Science Centre - the research project, 2010-2013.

analysis of changes of groups . The description consists of a set of graphs, each representing the blogosphere in a given time period (called "time slot"). The nodes of each graph represent active users (bloggers or identified users, who do not write blogs, but only add comments to existing ones) and the edges between them are weighted and directed, taking into account the numbers of comments written by a given user/blogger to the posts of bloggers in the given time period. It is assumed, that to add an edge, initiated by the commenting author to the posting author, it is necessary that the number of comments is higher than a given threshold value. In the experimental part of this paper, it is assumed that the existence of two comments in the given time period is sufficient.

The algorithm for predicting the states of the groups consists of the following steps:

Step 1. Identification of fugitive groups in the separate time periods
Step 2. Identification of group continuation – assigning transitions between groups in neighbouring time steps.
Step 3. Separation of the stable groups (lasting for at least three subsequent time steps).
Step 4. Identification of types of group changes. Assigning events from the list describing the change of the state of the group to the transitions

*A. Identification of fugitive groups*

On the basis of the data about comments written in given time periods, a set of snapshots of the network for each of the time slots is calculated.

For each obtained graph, the identification of fugitive groups is performed, using Clique Percolation Method (CPM) [16] algorithm, in a version for directed weighted graph. In the results, groups described by sets of their members in each time period are calculated. A similar method was described in [24].

*B. Identification of group continuation*

The next step is to associate groups that exist in the different time periods which are instances of the same groups in previous or next states of their existence. The algorithm identifies transitions between groups from the time periods t with groups from the time periods t+1, which are their successors. The decisions may be performed in two ways using two slightly different conditions:

Condition a: the group from the later time slot is the continuation of the group from the earlier time slot, if the modified Jaccard measure for these groups is higher or equal to the defined threshold value (set to 0.5 in the tests). The modified Jaccard measures are calculated for each formed pair of groups such that the first group is taken from the previous time slot, and the second – from the successive time slot. The modified Jaccard measure is expressed as a ratio of size of intersection of the pair of considered groups to the size of one of the groups from them (the larger value of such a ratio is considered as the modified Jaccard measure). For groups A and B (when both groups are not empty), the measure has the form:

$$MJ(A,B) = \max\left(\frac{|A \cap B|}{A}, \frac{|A \cap B|}{B}\right)$$

or otherwise $MJ(A,B) = 0$

Condition a: apart from the condition (a) the second condition also has to be fulfilled – a standard Jaccard measure is also calculated, and for it a low value of threshold is set (in the tests: equal to 0.01), to avoid the huge difference between the group sizes during the continuation. For the significant difference between their sizes, the small group has a very unimportant influence on the larger one, especially in cases when the groups overlap each other, so it is difficult to assume them being different instances of the same group.

*C. Separation of the stable groups*

In this step, the stable groups for subsequent analysis are distinguished. It is performed by rejecting groups which do not exist in the required number of the subsequent time slots (in the tests, it was assumed that groups have to last through at least three subsequent time slots). This method of the identification of stable groups was also applied by us in [24].

*D. Identification of the types of group changes*

Each transition between the states of the stable groups is associated with the attribute describing the kind of change of the group. The following kinds of group changes were defined:

1. **split** – takes place when there are many successor groups and the group is split into several parts, the group, that the transition comes to, cannot differ significantly (in tests, the threshold was set to 10 times) from the largest of successor groups. If it is the largest group, the transition is treated as simple transition - constancy or change size respectively.

2. **deletion** – group disintegrates into many successor groups and the successor group of this transition is significantly smaller than the largest group from successor groups (in tests, it is assumed that the successor group should be at least 10 times smaller than the largest successor group).

3. **merge** – transition is one among few, which create a group in the next time slot, the size of the former group cannot differ significantly (in tests, the threshold equal to 10 times was set) from the largest of predecessor group for the group that is created in the next time slot (if it is the largest group, the transition is treated as simple transition - constancy or change size respectively).

4. **addition** – the given transition is one among several which create a group in the next time slot, the origin group for this transition is significantly smaller (as in the case of the previous values, in tests the threshold value of size differences was set to 10 times) from the largest of origin groups (as in the case of the previous values, in tests the threshold value of size differences was set to 10 times).

5. **split_merge** – in this situation, in the same time, a split of the original group and the joining of many groups into successor groups took place, this transition is labelled as split_merge if the addition is not assigned earlier (we consider that the addition has higher priority).

6. **decay** – the total disintegration of the group, which does not exist in the next time slot.

7. **constancy** - simple transition without significant change of the group size (in the tests the acceptable size change for this case is set to 3 group members).

8. **change size** – simple transition with the change of the group size (in the tests, the change size for this case was larger than 3 members).

### III. GED: THE METHOD FOR GROUP EVOLUTION DISCOVERY

Recently Bródka at. al. in [4] have introduced a new approach to group changes identification called GED. They identified 7 types of changes

1. **continuing**, when groups in the consecutive time windows are identical or when groups differ only by few nodes and their size remains the same. Similar to **constancy.**

2. **growing**, when new nodes has joined to the group, making its size bigger than in the previous time window. A group can grow slightly as well as significantly, doubling or even tripling its size.

3. **shrinking**, when nodes has left the group, making its size smaller than in the previous time window. Like in case of growing, a group can shrink slightly as well as greatly.

4. **merging**, when a group consist of two or more groups from the previous time window. Merge might be (1) *equal*, which means the contribution of the groups in merged group is almost the same, or (2) unequal, when one of the groups has much greater contribution into the merged group. In second case merging might be similar to growing. Similar to **merge.**

5. **splitting** occurs, when a group splits into two or more groups in the next time window. Like in merging, we can distinguish two types of splitting: equal and unequal, which might be similar to shrinking. Similar to **split.**

6. **dissolving**, when a group ends its life and does not occur in the next time window. Similar to **decay**

7. **forming** of new group, which has not exist in the previous time window. In some cases, a group can be inactive over several timeframes, such case is treated as dissolving of the first group and forming again of the second one.

Next they introduced a new measure called *inclusion*. This measure allows to evaluate the inclusion of one group in another. Therefore, inclusion $I(G_1,G_2)$ of group $G_1$ in group $G_2$ is calculated as follows:

$$I(G_1,G_2) = \underbrace{\frac{|G_1 \cap G_2|}{|G_1|}}_{group\ quantity} \cdot \underbrace{\frac{\sum_{x \in (G_1 \cap G_2)} NI_{G_1}(x)}{\sum_{x \in (G_1)} NI_{G_1}(x)}}_{group\ quality}$$

$NI_{G_1}(x)$ is the value of node $x$ importance in group $G_1$. As a node importance, any metric which indicate member position within the community can be used, e.g. centrality degree, betweenness degree, page rank, social position etc. The second factor would have to be adapted accordingly to selected measures.

The *GED* method, used to discover group evolution, respects both the quantity and quality of the group members. The *quantity* is reflected by the first part of the *inclusion* measure, i.e. what portion of members from group $G_1$ is in group $G_2$, whereas the *quality* is expressed by the second part of the *inclusion* measure, namely what contribution of important members from group $G_1$ is in $G_2$. It provides a balance between the groups that contain many of the less important members and groups with only few but key members. Based on inclusion measure and group sizes the GED method is able to identify type of change between the pair of groups. For detailed explanation of GED see [4].

### IV. EXPERIMENTS

The aim of the experiments was to apply proposed methods dealing with identifying group changes on the same large data set and compare the obtained results.

*A. Data set*

The analysed data set contains data from the portal www.salon24.pl which consists of blogs, mainly political, but also have subjects from different domains. The data set consists of 26 722 users, 285 532 posts and 4 173 457 comments within the period 1.01.2008 - 31.03.2012. The analysed period was divided into time slots, each lasting 30 days. The slots overlap by 15 days and in the examined period there are 104 time slots.

We decided to remove edges with weights below 2 for two reasons. One of them was to eliminate some noise, because single comments can be random and should not be considered as an indicator of belonging to any group. The second reason was to reduce calculation time of extracting communities. After removing such edges, the number of nodes was equal to 15 578 (59.8 % of initial number of nodes) and the number of edges to 311 718 (47% of the initial number of edges). When we are considering the number of connections as the number of edges multiplied by their weights, then the removed edges constitute 8.42 % of such connections.

*B. Identification of groups in given periods*

For each time slot, the groups were obtained by the CPM (the CPMd from CFinder[1]) for different k in the range 3 to 5. Fig. 2 shows the average numbers of groups per slot aggregated by their size.

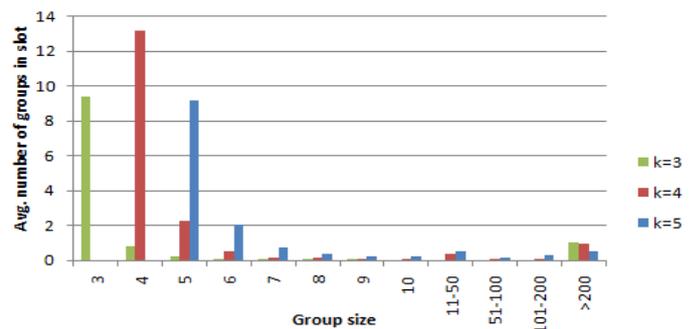

Fig. 2 Average number of groups in slot at given size for different k.

---

[1] http://www.cfinder.org/

## C. Transitions between stable groups

The axis x in Fig. 3 shows in percentage terms what part of one group transfers into another one. The diagram on this Fig. displays how many such transitions appear in the data.

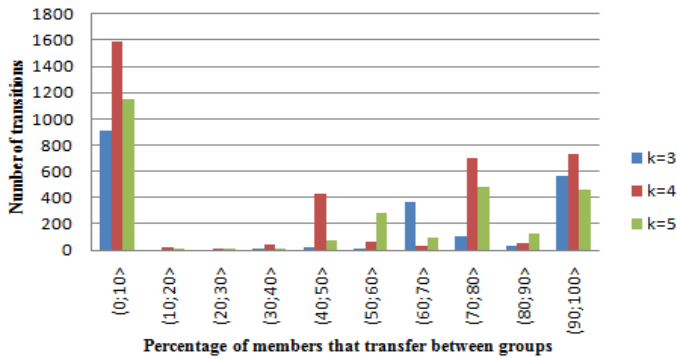

Fig. 1 Transition between groups

The first method of defining transitions between groups (based on modified Jaccard measure) has a threshold equal to 0.5. The part in Fig. 3 that shows transitions with a percentage lower that 50% describes states of group splitting: split, deletion and split_merge.

## D. Memberships of people to stable groups

Fig. 4 shows the numbers of users that belong to exactly one group. One can notice that with the increase of the k parameter the number of memberships to exactly one group decreases.

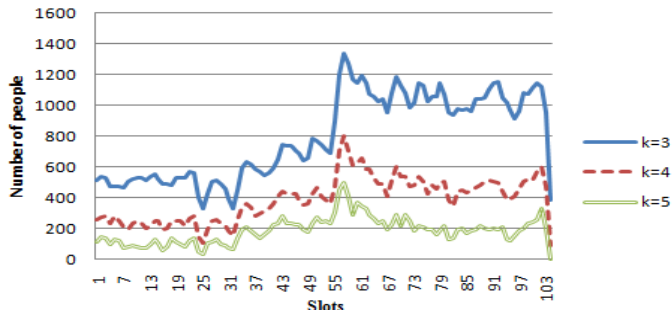

Fig. 2 Number of users that belong to 1 group

There is an interesting observation that the first half of the slots has a significantly lower number of users that belong to one group than in the second half of the slots. This change of behaviour takes place in the time slot number 55 (from 5 April 2010 to 5 May 2010) and it is a the result of the crash of the aircraft with Polish President and other officials near to Smolensk which occurred 10 April 2010. After this day, the frequency of posts and the traffic on the portal increased substantially, causing the formation of more groups.

TABLE 1 USERS THAT BELONG TO 2 OR 3 GROUPS

| k | membership to 2 groups | | membership to 3 groups | |
|---|---|---|---|---|
| | mean | std dev | mean | std dev |
| 3 | 11,92/30,71 | 4,7/10,32 | 0,5/1,54 | 0,67/2,08 |
| 4 | 22,08/53,29 | 7,75/14,94 | 3,46/9,37 | 2,57/4,31 |
| 5 | 18,23/39,15 | 7,64/12,46 | 3,75/9,92 | 3,03/5,71 |

Such behaviour appears also for a number of users that belong to exactly 2 or 3 groups, but the numbers are much lower - This is shown in Table 1. It seems that for each k parameter, the numbers of users that belong to a specified number of groups in the second half of the slots are about 2-3 times larger than in the first half of the slots.

## E. Overlapping other groups in time slots

The axis x in Fig. 7 shows what part of a group (in percentage terms) overlaps with other groups in the same time slot. The diagram presents how many pairs of groups with such defined ranges of overlapping percentage occur in the data set.

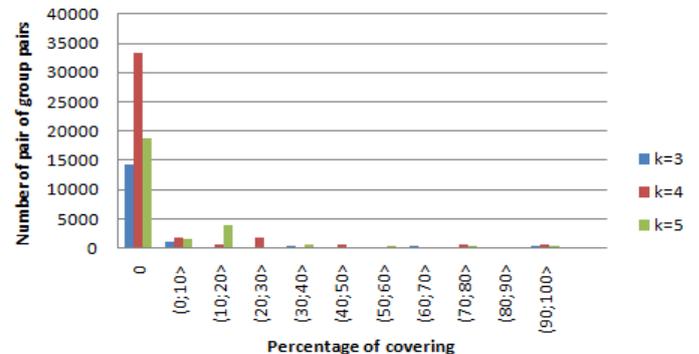

Fig. 3 Covering between groups in slot

## V. COMPARISON BEETWEEN SGCI AND GED

### A. The number of events

Table 2 and Table 3 show the numbers of given event types, identified respectively by SGCI and GED, for the whole considered time period (all slots) and for different k values of CPM algorithm. Some events considered in one systems are more detailed than in the other (merging in GED corresponds to merge and addition in SGCI, splitting in GED to split and deletion in SGCI and *change size* in SGCI to growing and shrinking in GED). Comparison considers these relations and shows appropriate matching of events in Table 4. The conclusion from Table 4 is that for merging, splitting and dissolution the SGCI discovers more event of such types, but the number of *change size* and constancy events extracted by the SGCI method is lower than from the GED method.

TABLE 2 NUMBER OF EVENTS EXTRACTED BY SGCI

| Event type in SGCI | k=3 | k=4 | k=5 |
|---|---|---|---|
| merge | 0 | 21 | 55 |
| addition | 953 | 1857 | 1283 |
| split | 0 | 46 | 35 |
| deletion | 834 | 1308 | 1006 |
| changeSize | 1 | 3 | 10 |
| constancy | 25 | 17 | 61 |
| decay | 157 | 73 | 97 |
| split_merge | 193 | 421 | 271 |

TABLE 3 NUMBER OF EVENTS EXTRACTED BY GED

| Event type in GED | k=3 | k=4 | k=5 |
|---|---|---|---|
| merging | 694 | 1177 | 964 |
| splitting | 726 | 1225 | 932 |
| growing | 66 | 91 | 72 |
| shrinking | 66 | 86 | 97 |
| continuation | 140 | 90 | 46 |
| dissolving | 86 | 54 | 62 |
| forming | 81 | 41 | 52 |

TABLE 4 COMPARISON THE QUANTITY OF EVENTS IN GED AND SGCI METHOD

| Event type [SGCI/GED] | k=3 | k=4 | k=5 |
|---|---|---|---|
| merge+addition/merging | 953/694 | 1878/1177 | 1338/964 |
| split+deletion/splitting | 834/726 | 1354/1225 | 1401/932 |
| decay/dissolution | 157/86 | 73/54 | 97/62 |
| changeSize/growing+shrinking | 1/132 | 3/177 | 10/169 |
| constancy/continuation | 25/140 | 17/90 | 61/46 |

In the Figs. 4-8 the quantities of transition events of given types identified by both algorithms in the respective time slots for the groups obtained for k =5 are displayed:

- **Merge**. Fig. 4 shows the amounts of transitions related to group merging. The SGCI discovered more events of this type.

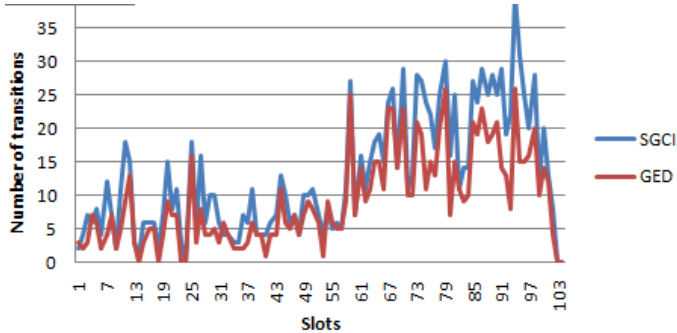

Fig. 4 SGCI:merge+addition/GED:merging

- **Splits**. Fig. 5 presents quantities of transitions related to group splitting. Also in this situation the SGCI method extracted more events than the GED method.

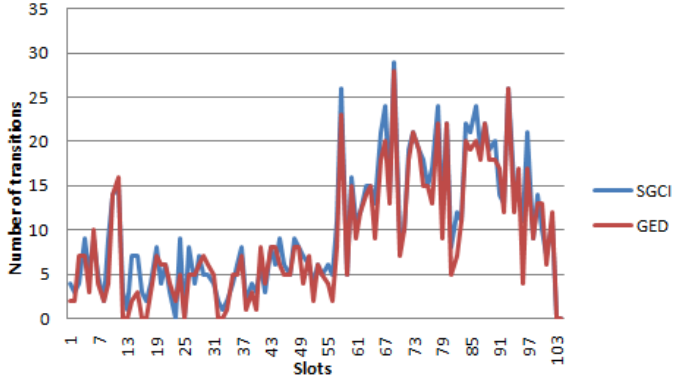

Fig. 5 SGCI:split+deletion/GED:splitting

- **Change size**. In Fig. 6 we can observe that the amount of events related to simple transitions with change of size for the GED method is greater than from the SGCI method.

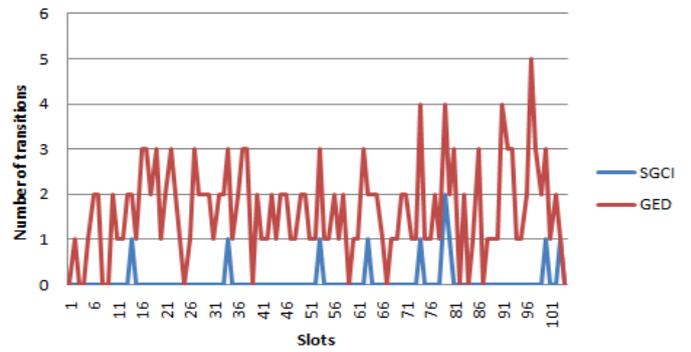

Fig. 6 SGCI:changeSize/GED:growing+shrinking

- **Constancy**. Fig. 7 describes transitions corresponding to simple transitions without change of size. The GED method has extracted more events of such type than the SGCI method.

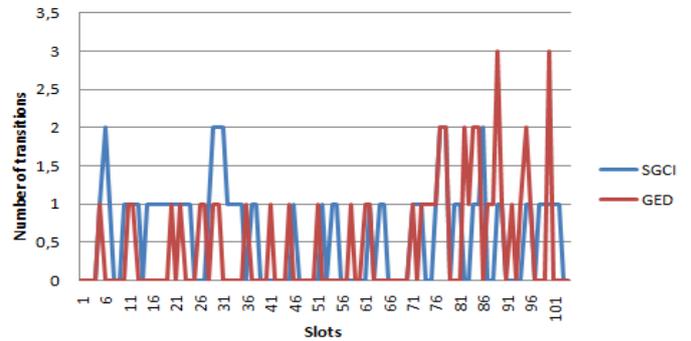

Fig. 7 SGCI:constancy/GED:continuation

- **Decay**. Fig. 8 presents the amounts of events that describe group dissolution. The SGCI method has discovered more of this type of event than the GED method.

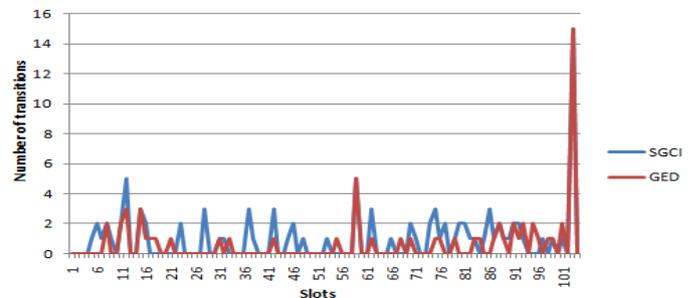

Fig. 8 SGCI:decay/GED:dissolution

*B. Events discovered by GED and not discovered by SGCI*

Table 5 shows the events that GED method has discovered and SGCI method has not. For forming event type the reason is clear - SGCI does not consider such a type of event. Apart from that, all other transitions were extracted by SGCI algorithm but were rejected as transitions from unstable groups. This is not surprising when we look into the details of both methods. The element of the modified Jaccard measure used by SGCI method is a part of the inclusion measure used by GED method. The GED method defines more strict conditions for transitions.

TABLE 5 EVENTS THAT GED HAS DISCOVERED AND SGCI HAS NOT

| Event type | k=3 | k=4 | k=5 |
|---|---|---|---|
| forming | 81 | 41 | 52 |
| rejected by SGCI as unstable transitions | 70 | 17 | 32 |

On the other hand, in the GED method there are more transitions only related to group continuation and change of their size during simple transition than in the SGCI algorithm. This fact can be explained that some transitions labelled as split_merge in the SGCI method are related to states: continuation, growing or shrinking from the GED method. In the above figures the state split_merge have not been included into comparison with any of states from the GED method. But the number of transitions labelled as split_merge is quite large.

*C. Events discovered by SGCI and not discovered by GED*

The total number of events found by both methods vary from 2379 to 3848. As the total number of events found by the methods grows, the number of events found by the SGCI method and omitted by the GED method increases likewise.

The reason why GED did not find those events are the method parameters alpha and beta (See [4] for detailed explanation). All events omitted by the GED method have both inclusions below 50%, which was the value for alpha and beta thresholds.

To prove this, the GED method was calculated again with thresholds equal to 10%. This time, only a few events (less than 10) found by the SGCI method were omitted by the GED method. Both inclusions of the omitted events are below 10%. In the results of the GED computation some decay events, found by the SGCI method, were missing as well. This is caused by different requirements (e.g. threshold values) of both methods for dissolve (decay) events.

TABLE 6 EVENTS THAT GED HAS DISCOVERED AND SGCI HAS NOT

| | Total no. of events found | | No. of events which second method did not find | | Total no. of distinct events found by both methods |
|---|---|---|---|---|---|
| k | GED | SGCI | GED | SGCI | GED & SGCI |
| 3 | 1859 | 2163 | 216 | 520 | 2379 |
| 4 | 2764 | 3746 | 102 | 1084 | 3848 |
| 5 | 2225 | 2818 | 128 | 721 | 2946 |

## VI. CONCLUSION

Two separate methods for identifying changes in social group evolution have been presented and considered in the paper: (i) new method for stable group identification (SGCI) as well as (ii) existing method for group evolution discovery (GED). Each method defines its own types of group changes (most events are similar in both methods) and measures (modified Jaccard measure in SGCI and inclusion in GED). The important difference was also that the SGCI was operated only on identifiably stable groups, where the duration of the group was at least equal to the given period, whereas GED worked on all groups.

Both methods were applied to the data obtained from the Polish blogosphere. The main part of the experiments concerns the comparison of results obtained with both methods in terms of events detected by the first method but not by the other and vice versa. The results are similar (only few events found by one method were omitted by the other).

Further work will focus on adding the prediction mechanism to predict the future changes of the groups on the basis of their current and previous states.